# ENERGY AWARE ROUTING WITH COMPUTATIONAL OFFLOADING FOR WIRELESS SENSOR NETWORKS

Adam Barker and Martin Swany

Department of Intelligent Systems Engineering,
Indiana University, Bloomington, Indiana, USA

## ABSTRACT

*Wireless sensor networks (WSN) are characterized by a network of small, battery powered devices, operating remotely with no pre-existing infrastructure. The unique structure of WSN allow for novel approaches to data reduction and energy preservation. This paper presents a modification to the existing Q-routing protocol by providing an alternate action of performing sensor data reduction in place thereby reducing energy consumption, bandwidth usage, and message transmission time. The algorithm is further modified to include an energy factor which increases the cost of forwarding as energy reserves deplete. This encourages the network to conserve energy in favor of network preservation when energy reserves are low. Our experimental results show that this approach can, in periods of high network traffic, simultaneously reduce bandwidth, conserve energy, and maintain low message transition times.*

## 1. INTRODUCTION

A wireless sensor network (WSN) is characterized by a network of small, low power devices, operating remotely with no pre-existing infrastructure, and little to no human intervention or central management. They typically rely on on-board energy storage such as batteries and can be required to operate for months or years at a time. Whereas many WSN collect sensor information to be relayed to a central location, some instances, particularly those for use in military and first-responder applications operate in a peer-to-peer paradigm in which a sensor node can also be a consumer of sensor information from other nodes. This peer-to-peer concept implies any node can be both a source and destination making routing paths between source and destination dynamic. Because WSN operate without fixed infrastructure, they must handle their own routing and can form mesh networks where sensor information can traverse several hops across other intermediate sensor nodes to reach their destination. The lack of fixed infrastructure also leads to limited access to bandwidth. Typical WSN operate in the tens to hundreds of kilobits per second such as seen in IEEE 802.15.4 or Semtech's LoRa protocol. Every bit being transmitted comes at a cost of not only energy to transmit over the wireless link, but also a cost of time to get critical data where it needs to be. As data sets continue to increase in size, the hundreds of kilobits per second data rates will continue to form a bottleneck. The peer-to-peer pattern coupled with a mesh topology can offer some unique opportunities to help reduce bandwidth consumption, reduce the time data needs to travel in the network, and preserve scarce energy reserves.

Until recently, wireless sensor networks have seen limited use, however the concept of the Internet of Things (IoT) has begun to gain popularity putting WSNs at the center of focus as one of the technologies needed to enable the IoT. The IoT promises networks of millions to billions of small devices connecting everything from wearable technologies to autonomous drones. For this scale of technology to be realized, energy, bandwidth, and data delivery time become critical aspects that need to be properly managed. The unique structure of WSNs allow for novel approaches to data reduction and energy preservation.

This paper presents a concept to assist in management of these factors by re-examining existing routing concepts and improving upon them for the unique use case of peer-to-peer mesh WSNs. This concept is implemented through a series of adaptions to the Q-routing protocol we collectively refer to as energy aware Q-routing with computational offloading (EAQCO.) Typical WSN routing algorithms only route messages based on least-cost routes without considering in-situ computation options to reduce the message data prior to forwarding. The EAQCO concept optimizes trade-offs between energy-expensive message passing and time-critical computational offloading in an effort to deliver usable information to a destination node within a WSN. To the best of our knowledge the EAQCO concept is the only research that manages the trade-offs of delay, energy consumption, and bandwidth in a WSN. The results of our experiments show that EAQCO is a viable concept that can minimize time to deliver processed data while minimizing bandwidth used, particularly in networks with high data traffic.

The remainder of this paper will be divided into sections, starting with section 2, which describes the background of WSN routing research. Section 3 will then present a novel approach to address the major concerns of time, energy, and bandwidth, implemented through EAQCO. Finally, section 4 will present results of experiments performed on a wired mesh network of small IoT type devices implementing the EAQCO algorithm.

## 2. BACKGROUND

Wireless sensor network routing has been the subject of much research over the past 30 years. With the implementation of 5G technologies, specifically concepts involving Multi-Access Edge Computing (MEC), and increased interest in the Internet of Things (IoT), research in WSN routing has been on the rise. Research in this area can be grouped into 3 general focus areas: energy efficiency, computational offloading, and traffic/congestion management.

### 2.1. Energy Efficiency

Because WSN operate from energy storage devices such as batteries, maintaining the longevity of the network is the focus of this area. Notable examples of current and past research focused on energy efficient routing include [1], [2], [3], [4], and [5]. Much of the earlier research, such as [1] and variations such as [2], focus on clustering of nodes and aggregation by a selected (in the case of LEACH, randomly selected) cluster head before forwarding data to a fixed end point. Other variations of LEACH allow nodes to enter a low power sleep mode to further conserve energy when they are not a cluster head or transmitting data, such is in [3]. The concept of clustering is very viable for a densely populated sensor network, but in a sparse, widely distributed, network where nodes may only have 2-3 neighbours by which to route data through, the number of cluster heads may come close to the number of nodes in the network. Additionally, clustering relies on sharing the cost of transmission. If the nodes are heterogeneous, particularly regarding energy storage capacity, equally sharing the cost of being a cluster head may not be the most productive approach.

Other novel approaches such as [4] formulate the energy maximization problem as linear program and use multi-commodity flow algorithms to solve for minimum cost, maximum flow where the cost is measured in remaining energy storage and flows are data transmissions. Maximum flow algorithms are widely available and relatively efficient to compute giving an optimum solution very quickly. However, to formulate the affine function across the entire network requires constant updates of the exact network topology including knowing the energy storage of each node at any given time. For small to medium networks with constant transmissions spanning the entire network this is a viable solution, but again for sparse heterogeneous networks, flooding the entire network with all nodes' energy storage state may require significant additional bandwidth and energy to ensure all nodes receive the most up-to-date topology information.

More recent works have examined more advanced approaches such as the use of genetic algorithms in [5]. Their work is novel, however the requirement for a middle layer, similar to the notion of a cluster head, and the construction of their genetic algorithm, require an existing model of the network. In many cases of WSN deployment the exact structure of the network may not be known and therefore no model can be built *a priori*.

## 2.2. Computational Offloading

Computational offloading is the process of moving the task of processing raw data to a node or set of nodes physically separated from the sensing node. Much work has been done in this area and has developed into core businesses such as Amazon Web Services and Google's Cloud Services. With the increase in the volume of data, sensor nodes with typically reduced computational capacity need to move local computation to a central location for processing, however larger cloud services require high bandwidth, typically orders of magnitude more than WSNs are able to provide. For IoT devices, the majority of the research has focused on building relatively lightweight algorithms and processes that adapt cloud services for mobile networks outlined in work such as [6] and [7]. With the advent of 5G networks, research into computational offloading has turned toward the MEC. The research concept for MEC typically focuses on a centralized algorithm to manage the decision process between multiple nodes within a network and the edge nodes and servers that can perform the offloading. One recent example, [8], uses a model of the network, the self-reported transmission time, and estimated energy consumption of each node to construct a game theory algorithm that settles into a Nash equilibrium for the optimum strategy for all nodes and all tasks in the network.

Each of these approaches are novel and have merit given the use case presented, however the current research is limited in two ways. Firstly, the offloading algorithms require a network model for which to optimize the decision process against, and secondly the algorithm is centrally planned and managed in which each individual node receives the offloading decision from a central node, or base station. In a purely peer-to-peer *ad hoc* network the network model may not be available and distributed decision making would need to take the place of central management. Additionally, most of the offloading models involve arbitrarily complex functions that would need to be offloaded from a multi-function device such as a wireless handset, however WSNs typically involve a single function, or limited functions, on devices that run limited scope computational algorithms. These devices may not benefit from a complicated model that hands off functions via passing of virtual machine objects as is common research thread for MEC.

## 2.3. Traffic/Congestion Management

Perhaps the most widely researched area in WSN routing is that of traffic management. Data gathered via WSN is typically time sensitive, in the case of real-time applications measured in the microseconds, and therefore optimizing the fastest route to the destination is of highest concern. Routing in a WSN, like other data networks, is often resolved using shortest path algorithms such as ad hoc on-demand distance vector (AODV), optimized link state routing (OLSR) and dynamic source routing (DSV). Variations of these algorithms exist in research in an effort to optimize energy usage and node mobility which are both key features of WSN. The limitation of these algorithms is they rely on information that is gathered in an instant in time, either as needed, or at some time in the past. Due to the dynamic nature of WSN, these protocols do not allow for prediction or learning of the dynamics of the network.

To address variability in wired networks the authors in [9] developed a reinforcement learning (RL) approach to compare with shortest-path-first algorithms used in most networks. Using their example, a message is required to be sent from source $x$ to destination $d$ via its neighbour $y$. Their approach was a simple variation of the RL concept called Q-learning using the formula:

$$Q_x(\bar{y}, d) = Q_{\bar{y}}(\bar{z}, d) + q_y \tag{1}$$

where $Q_y(\bar{z}, d)$ is $y$'s estimate of the remainder of the message's journey to $d$ after it leaves $y$ and $q_y$ is $x$'s estimate to get to $y$. Whenever $x$ receives a reply from $y$ it includes $y$'s estimates to get to each destination, known as $y$'s Q-table. Using $y$'s Q-table, $x$ updates its estimate to $d$ using the update rule:

$$Q_x(\bar{y}, d) = Q_x(\bar{y}, d)^{old} + \eta\left(Q_{\bar{y}}(\bar{z}, d) + q_y - Q_x(\bar{y}, d)^{old}\right) \tag{2}$$

Where $\eta$ is known as the learning rate. Neighbour $y$ may have several routes to $d$ therefore $x$ selects $Q_y(\bar{z}, d)$ using:

$$Q_{\bar{y}}(\bar{z}, d) = \min_{z \in neighbors\ of\ \bar{y}} Q_{\bar{y}}(\bar{z}, d) \tag{3}$$

Each node maintains its own Q-table using Equations 1, 2, and 3 therefore Q-routing is a distributed learning process.

Many variations of the Q-routing process have been researched to include [10] which adds a confidence factor and backward exploration to address the probability a node in the ad hoc network may drop out periodically. Additional variations include [11] which incorporates a separate learning phase to develop quality of service metrics and [12] which combines any cast routing with the learning capabilities of Q-routing.

## 2.4. Other Related Work

Our work retains elements of energy efficiency, computational offloading, and congestion management to build a novel approach to routing based off of the Q-routing algorithm from [9] and is detailed in section 3. Other approaches have examined similar aspects, such as in [13] and [14].

In [13], researchers develop a computational offloading scheme for internet of vehicles (IoV) applications. Their work utilizes a genetic algorithm to optimize the offloading of compute tasks to edge servers. Additionally, they include the option to route offloading tasks through adjacent nodes that may be in range of alternate edge servers. While their approach is novel, their multi-objective optimization approach requires global knowledge of every compute task in the local network. In a peer-to-peer WSN such as we outlined above, global knowledge of required computing tasks is not available to individual nodes without imposing a significant cost of communication and time to gather global data at a designated head node. Therefore, we do not believe their formulation is applicable to our intended use case.

In [14], researchers examine industrial IoT (IIoT) applications of multi-hop computational offloading and develop an algorithm using a game theory approach. Their approach is distributed as each IIoT node makes its decision to offload computation independently of the next node. The decision process works as a game giving each node the option to determine if it is more advantageous to compute locally versus remotely. The game ends when all nodes reach a Nash equilibrium. Their algorithm allows for multi-hop routing between nodes and edge compute nodes making it a routing optimization problem. While their efforts are similar to our use case, their model is limited to specific edge compute nodes and doesn't allow for transfer of compute capabilities to adjacent nodes. In a peer-to-peer WSN there is no designated edge compute node and any determination of computational offloading benefit is made at each node based on its local ability to perform compute functions.

The next section discusses our approach to the unique problem of peer-to-peer WSN and addresses the issues that research to date has not taken into account.

## 3. DISCUSSION

In an effort to improve the efficacy of Q-routing specifically in a WSN environment while maintaining awareness of limited energy storage capacity and utilizing the computational power latent within the network itself, we present the concept of energy aware Q-routing with computational offloading (EAQCO.) The EAQCO concept utilizes the simplicity and distributed capabilities of the Q-routing algorithm and adds in additional decision logic that determines if it is more feasible to perform data reduction in-place or forward raw data. The primary metric to determine optimal route selection is time, however the energy-awareness component adds an additional factor that increases the cost as energy reserves become depleted.

### 3.1. Computational Offloading

The EAQCO concept begins with the basic Q-routing algorithm. For an uninitialized network, a series of ping messages carrying a timestamp are sent from a node to all of the node's neighbours. Each neighbour immediately responds to the ping with time it took to receive and process the ping. These responses form the basis of the node's Q-table. In addition to the time to receive the message, the node's neighbours include in their response, a copy of their own Q-tables from previous iterations of ping messages sent to their neighbours. These neighbour Q-tables contain the best estimates of time to send messages to their neighbours. As the ping messages and responses continue for a few iterations the entire network is mapped out with estimates of time to send data between any two nodes in the network.

When a node receives its neighbour's Q-tables it adds the time it takes to send data to its immediate neighbours and the time recorded in the neighbour's Q-tables to build an estimate of the total time it takes to send a message to each destination covered in the neighbour's Q-table.
The sending of pings is only needed to establish the initial topology of the network and can be completed with a maximum number of pings equal to the longest route in the network. Once the topology is established and messages containing data are sent throughout the network, each node that receives a data message acknowledges the receipt of the data with a response identical to the receipt of a ping. Therefore, each node's Q-tables are continually updated as long as data is flowing in the network. For sparse communications and/or to verify if various nodes are still reachable, pings can be sent on a periodic basis to ensure the Q-tables remain up to date. The authors in [9] noted that early versions of their Q-routing algorithm settled into an optimal route quickly but were unable to recognize when an alternate route was available. This concept, known as exploitation versus exploration, is a well-researched issue within the field of reinforcement learning. To address the issue a parameter, ε, known as the exploration factor, is included in the Q-learning algorithm. By adding in ε, instead of choosing the action with the highest Q-value, or in the case of Q-routing, the shortest time, there is a probability, ε, that the algorithm will choose a random route to explore. Values for ε are typically 0.1 - 0.5. Other factors such as those proposed in [10] could be added to the Q-routing algorithm as needed to enhance the performance, but they would not disrupt the computational offloading capabilities described below.

So far, this process described above is no different from standard Q-routing, however, EAQCO adds an additional step that is determined in parallel to message routing. When a data message is first made available for forwarding, and every time a message is received by a node, there is a decision process to forward or perform data reduction (computation) in place. If data reduction is selected, the message is placed in the node's computation queue and processed in the order it was received. The cost of computing in place is measured in time by determining the length of the processing queue and is updated using the same update rules for determining new Q-values shown in Equations 1, 2, and 3. This process is built into the Q-routing algorithm process at the point where the optimal route is selected.

With no loss to generality, a simple example is used to explain the decision process. Figure 1 shows an extremely simple network of just 3 nodes.

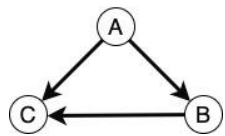

Figure 1. Simplified Example Network

Assuming Node A had a requirement to send data to Node C there are two paths it could take: A to C directly or through B. The Q-routing algorithm, given in [9], would determine the optimal route by choosing the minimum time of the 2 routes:

$$Q_A(c) = min\{ Q_x(\bar{c}, c). Q_x(\bar{b}, c)\} \qquad (4)$$

However, with the computational offloading step, a third option exists by including the Q-value for performing data reduction in place. The set of actions that can be performed by A are then:

Table 1. Actions available to A

| Action | Reward |
|---|---|
| Fwd to C | $Q_x(\bar{c}, c)$ |
| Fwd to B | $Q_x(b, c)$ |
| Compute Locally | $Q_x(compute)$ |

Therefore Equation 4 is now:

$$Q_A(c) = min\{ Q_x(\bar{c},c), Q_x(\bar{b},c), Q_x(\overline{compute})\} \quad (5)$$

If, in this simple case, the data could be reduced and sent to C in less time than it would take to send the raw data to C, A would choose to perform data reduction before forwarding the resultant information on to C. Because of the nature of WSNs, most of the data is redundant and can be repetitive, however without some data reduction step, it cannot be known what is useful data, therefore the default is to forward all data and let the end user determine what to keep. The computational offloading step performs the pre-defined reduction before the data reaches the end destination. The addition of the compute action is also included in B's action set. If A chose to forward to B, B now has two available actions: continue forwarding to C or compute in place, therefore the entire tree of actions for moving data from A to C is expanded with the inclusion of the option to compute locally at every node.

As an example, if a WSN of seismic sensors was monitoring the vibrations at a specified location they might be capturing data several hundred times a second. Once there is enough data it is forwarded on to the destination to determine when there is a significant change. Ultimately what is needed is the time and magnitude of the change point. This could be captured in a tuple of 2 32-bit numbers (time and magnitude), representing a total of 64 bits of information, however, the dataset needed to determine the change point may contain several thousand samples of 64-bit tuples. Therefore, computation can result in a 1000-fold reduction in bandwidth. Because WSN are typically made up of small computationally constrained devices, determining the change point may take a significant portion of the originating node's computational capabilities and could not be performed for every dataset without incurring significant delay. As the datasets become more complex and the computation more intensive, such as the case with depth mapping full motion video or IQ-processing for software defined radios, the cost of computation continues to increase and therefore efficiency benefits from computation being distributed throughout the network.

Because the EAQCO is distributed and the forwarding versus computation decision process is happening at each node, the decision to compute locally is done at each node and with each compute decision, the data set is slowly being reduced focusing on optimizing overall data propagation time. Each decision to compute has the added benefit of reducing the overall bandwidth needed to transmit the streaming data which reduces the overall energy needed to transmit. This detail will be revisited in a follow-on section.

Each iteration of a compute task updates the Q-value for local computations just as each forwarded message updates the Q-value for the time to forward a message to the next neighbour. As the process continues, each node builds its Q-table of forwarding routes and comparative computation costs. This is the Q-table that is returned after each data message is sent or a ping is received. Therefore, each node that receives a neighbour's Q-table is receiving their neighbour's decision to either forward a message or compute locally. This indicates if a node decides to forward a message to a neighbour because its Q-value is lower than the other options, that node does not know whether their neighbour will forward that message on or perform data reduction in place.

There are limitations to this approach, namely that the data and the resultant computation must be severable. If computation of one subset of data requires the entire dataset, this approach will not be feasible, however most streaming applications such as video, audio, and IQ-data are often severable and can be computed in slices as needed.

High level pseudo code of ping and data sending process with computational offloading is shown in figure 2.

```
Algorithm 1 Q-routing with computational offloading
Initialize the algorithm:
Initialize Q-value for local computation
Send pings to map network
if Received neighbors Q-tables then
    if New destination received then
        Update new entry in Q-table
    if Existing route received then
        Select minimum cost route
        Perform Q-learning to update Q-value
if Data Packet Received then
    if Destination Exists in Local Q-table then
        Determine optimal Q-value from table
        if optimal decision is compute locally then
            Add packet to computation queue.
            Measure queue length.
            Perform Q-learning to update Q-value for local
            computation
if Ping received then
    Send response including one-way time to sender and local
    Q-table
```

Figure 2. Pseudo code of EAQCO process

## 3.2. Energy Awareness

The timeliness of sending data within a WSN is of utmost importance for time critical applications such as military, public safety, and high-mobility platforms like self-driving automobiles. However, a unique feature of most WSNs is a limited energy storage capacity. They are typically battery operated and at times intended to operate for several months or even years without human intervention.  The scarce resource of energy presents another set of concerns for WSN.  To extend the battery life of wireless sensors and other energy constrained devices, many studies have researched the energy consumption profiles to quantify the subsystems with the largest energy consumption profile.  One example in [15] found that a mobile phone's Wifi or GSM module can consume more than 8 times the energy of the CPU.

Determining the energy used to transmit wireless messages is the subject of much research, however generally, energy is consumed in two parts: energy used to transmit radio frequency (RF) signals, and energy used to receive RF signals.  Typical wireless packet-sending protocols start with the transmission of a message, once it is received an acknowledgement is returned to confirm message receipt.  If no acknowledgement is received within a predetermined timeframe the message originator retransmits the message.  This process continues until the message is acknowledged or a retry threshold is met.  Determining the amount of total energy consumed during this transmit/acknowledge process is a stochastic process based on the characteristics of multiple wireless parameters.  Researchers in [16] examined this process and developed a simplified model that estimates the expected energy consumed.  Using the energy consumption model developed in [16], a formulation was developed to determine energy consumption based on data message size and includes factors for transceiver hardware characteristics and wireless link parameters.  The formulation developed is shown in Equation 6 for a data message of size $x$ bits and a data rate of $r$:

$$E[e_t] \leq \frac{1}{p^2}\left((P_{xmtr}) \times \frac{x_{data\_packet}}{r_{data\_rate}}\right) + \frac{1}{p}\left(\frac{P_{rcvr}}{r_{ACK\_rate}} \times x_{ACK\_packet}\right) \quad (6)$$

where $E[e_t]$ is the expected energy used to transmit a message, $p$ is the message reception probability, which is a function of the characteristics of the wireless link, $P_{xmtr}$ is the wireless radio transmit power (including losses due to transmitter inefficiencies), and $P_{rcvr}$ is the wireless radio receiver power consumption. The inequality is the upper bound considering message retransmits due to signal loss and applies as long as the number of retransmits is less than $\infty$.

From this formulation the cost of transmitting per bit over a particular link can be estimated giving a factor for transmitting future messages over a wireless link given the current energy storage of the wireless sensor device. This factor, we designate as the energy factor (*ef*), is expressed as an exponential function, the factors of the exponential are determined by the specific type of wireless link used as expressed in Equation 6, however the variation used in the proceeding experiments assumed a low data rate with a transmission power of milliwatts to one watt such as used in the LoRaWAN™ wireless link [17]. An exponential function was chosen, because we desire the *ef* to have minimal impact when there is minimal energy reserves used, such as below 50% energy capacity level, but asymptotically approach a factor of 10 as energy used is above 50%.

Using this configuration, the energy factor formulation becomes:

$$ef = 0.001 \times 10^{4 \times (batt\_used)} \times msn\_remain \quad (7)$$

where *batt_used* is the fraction of battery capacity consumed and can be retrieved from the on-board battery monitoring circuitry. The term *msn_remain* is an added factor to account for a WSN's potential for a predefined operating time, which is often the case in military applications. The *msn_remain* is the fraction of total estimated mission time remaining and devalues the battery capacity factor as the mission time gets closer to its expected completion point. For operations where the WSN is needed to operate indefinitely, this factor can be eliminated.

The energy factor is computed with every update to the Q-routing algorithm and essentially wraps the computed Q-value for a given forwarding route in the exponential function. As the energy factor increases, the overall Q-value for any given forwarding route increases as well, therefore as the energy storage reserves are depleted the advantage of computing in place increases even as the compute queue increases. The overall effect is data flow rate within the network decreases as energy reserves deplete, shifting the priority from optimizing the information delivery time to overall network preservation.

Given Equations 1, 2, 3, 5, and 7, and a requirement to send information from *x* to *d*, the final formulation of the EAQCO algorithm becomes:

$$Q_x(y,d) = Q_x(y,d)^{old} + \eta \left(ef(t_{x \to y}) + \min_{z \in neighbors\ of\ y} Q_y(z,d) - Q_x(y,d)^{old}\right) \quad (8)$$

$$A_x(d) = min\{Q_x(y,d), Q_x(\overline{compute}) \forall y \in Y\} \quad (9)$$

Where $Q_x(y,d)$ is *x*'s value of forwarding a message to *d* through neighbour *y*. Greek letter $\eta$ is the learning rate, *ef* is the energy factor from Equation 7, and $t_{x \to y}$ is the estimated time to send a message from *x* to *y*. The element $Q_y(z,d)$ is the Q-table forwarded from all of x's neighbours, including any of *y*'s calculation of $Q_y(\overline{compute})$ which is their estimation to compute locally.

Therefore, *x* selects the min cost action, $A_x(d)$, between either forwarding to any of its neighbours in the set of all neighbours *Y* or computing locally as shown in Equation 9.

Equations 8 and 9 form the basis for the EAQCO algorithm. The addition of the energy factor in Equation 8, bias the forwarding of messages based on the stored energy reserves of a given node and the inclusion of the action to compute locally to reduce energy consumption and bandwidth usage are the primary contributions of the EAQCO algorithm. Using these Equations, several experiments were run as is described in the following section.

## 4. EXPERIMENTS

To test the application of EAQCO, a physical network of Beaglebone® Blacks [18] was built, where each node contained at least two network devices and could perform as a router for network traffic while simultaneously performing sensor and compute functions. The network was instantiated using copper Ethernet as the communication link as physical limitations of the lab did not allow for the use of a wireless link. The algorithm was built in Python3 using the Sockets API with a UDP transport protocol for all messages. Sensor data was simulated by randomly generating data sets of 100 16-byte floating point numbers with an associated 16-byte timestamp. The overall message size was 2360-2400 bytes including header information. The configuration of the network is shown in figure 3 with each node corresponding to a single Beaglebone® Black.

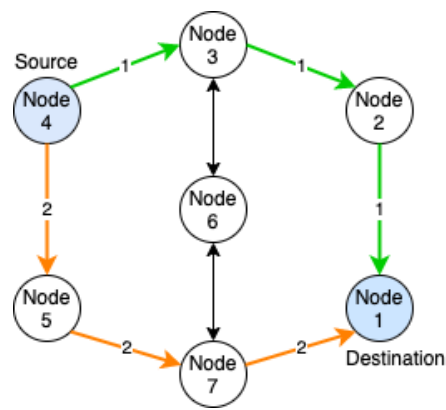

Figure 3. Beaglebone Test Network

### 4.1. Test Setup

Three scenarios were tested while varying different parameters as shown in the results section:

- Static Routing
- Q-routing with Computational Offloading
- Q-routing with Computational Offloading and Energy Awareness

For the Q-routing cases, each node is running an instance of the EAQCO algorithm and is generating and receiving corresponding ping and data messages, however data was only gathered between Node 4 (source) and Node 1 (destination) to reduce the size of the data set. For the static case, each node routes messages along a predefined static route, however this route could have been developed using any standard *ad hoc* routing algorithm such as AODV or OLSR. For all cases, all nodes, except for source and destination, generate random data destined for any of the randomly selected nodes to ensure the network is sufficiently loaded, commensurate with the specific phase of testing.

In each scenario, Node 4 streams a series of 750 messages containing payloads of 2360 bytes to the destination, Node 1. In the static scenario, the messages follow route 1 as designated in figure 3. In the alternate scenarios, both route 1 and route 2 are utilized and occasionally the algorithm will explore sending data via a path that traverses Node 6. All three scenarios were run multiple times in configurations designated low and high where in the low configuration the intermediate nodes are generating random data messages approximately once per second and in the high configuration messages are generated sequentially as fast as possible. The high rate typically corresponded to an effective data rate of approximately 1.5 Mbps/node. The network loading corresponded to an approximate 10-fold increase in message transmit times between low and high.

Each test scenario was run 10 times and the results were averaged over across the runs. Once a baseline was established for the three scenarios, additional tests were performed varying the learning rate and ε parameters. The results of the tests are shown in the next section.

### 4.2. Results

Data was collected to look at the primary metric of overall effect on message processing times. Processing time, in the context of the tests, is defined as the time a message takes to transition from source to destination, including time to perform the necessary data reduction/computation. For the static routing case all data reduction/computation takes place at the destination node after the message traverses the predefined route. For the Q-routing cases, computation can take place anywhere in the network as determined by the decision process of the routing algorithm. The results of the 10 trials of each low and high data rates are show in figure 4.

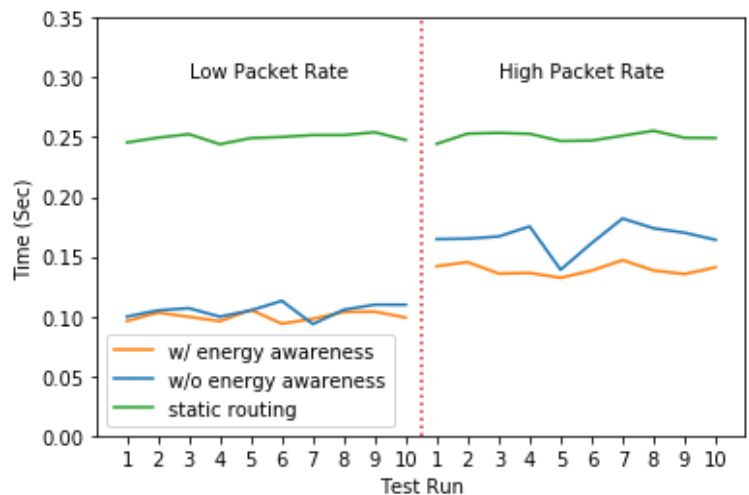

Figure 4. Mean total processing time per message

It can be seen in the results that Q-routing with computational offloading results in a significant reduction in total message processing time. For the low message rate case, energy awareness does not appear to affect the mean time significantly, however in the high message rate case there is a marked difference as the decision process tends to skew towards computation in place rather than forwarding.

In addition to timing, data was examined to determine the overall energy usage of the message stream's traversal from Node 4 to Node 1. The mean energy usage of the 10 trials at each the low and high configurations was determined using the energy model from Equation 6, assuming no retransmits, and hardware profile equivalent to LoRaWAN™ to determine energy usage per byte transmitted. Each trial used a learning rate η of 0.1 and an ε of 0.1. Total energy used was then calculated using:

$$E_{total} = s \times h \times p$$

where *s* is the size of the message in bytes, *h* is the number of hops, and *p* is the total number of messages in each stream. In the static routing case all factors for energy are constant, which results in a linear result across all test runs. For both cases of computational offloading the number of hops varied depending on the optimal decision made at each node. The results are shown in Figure 5.

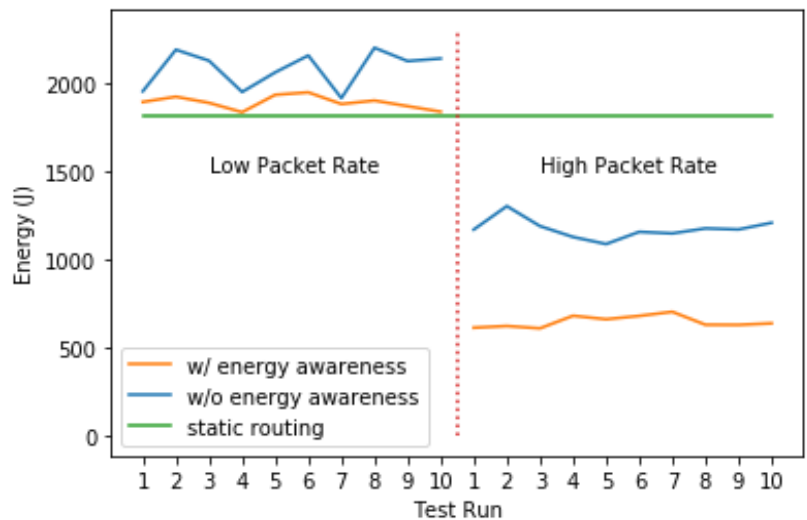

Figure 5. Mean total energy used to transmit data stream

For the low rate, both versions of computational offloading result in higher energy usage compared to the static baseline. There are three factors contributing to this effect. First the Q-routing algorithm performs exploration as noted previously, therefore, randomly, with a probability ε the algorithm may choose a more energy costly route such as traversing the path through Node 6. Secondly the primary factor for route optimization is time rather than number of hops, therefore, due to some congestion at Node 3 the algorithm may find the optimal route is through Node 6, which minimizes time rather than energy. This is highlighted in the difference between the case with energy awareness and the case without. As the energy storage capacity is depleted, the algorithm favours lower energy routes; fewer hops over more hops. Thirdly, in the low data rate case, the cost to compute locally does not compare favourably to the relatively short message transit times therefore the algorithm is forwarding messages more than computing locally. Conversely, examining the high message rate results show the effect of congestion in the network as the algorithm is deciding to compute locally rather than forward at a higher rate. The greater difference between the two cases of with and without energy awareness is a result of the increase in message processing at all nodes in network. The more messages that are processed the more the energy storage reserves are depleted and the more favourable the decision compute becomes. This can be further highlighted in figure 6.

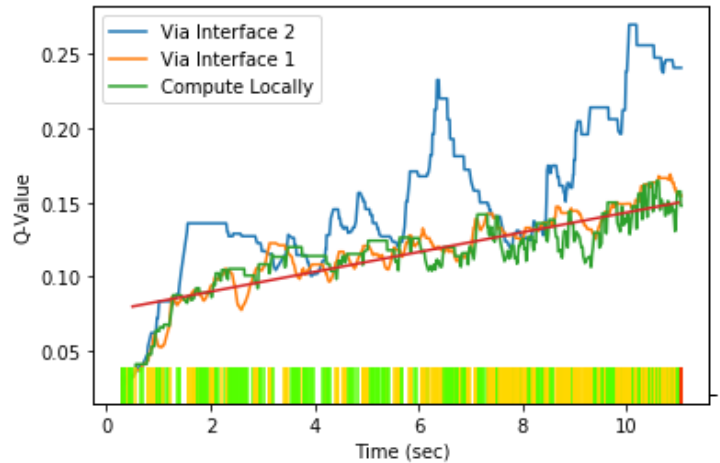

Figure 6. Q-value of action space over time

Figure 6 shows the Q-values over time from Node 4 in a randomly selected run. The red line highlights the upward slope as energy reserves are depleted. The bottom tick marks show action selections where green corresponds to forwarding and yellow corresponds to computing locally. It is also worth noting that the Q-value for computing locally also increases over time. This is due to a higher rate of decisions to compute locally resulting in longer compute queues. The long-term effect of this is the entire action space increases as energy reserves are depleted and, although it becomes more favourable to compute in place, no one action becomes completely dominant.

Once initial results were examined with a set learning rate and set exploration factor, additional experiments were run varying either factor to examine the effects. For this set of experiments, only the Q-routing with computational offloading and energy awareness algorithm was utilized. The experiments were run using the variable data rate loading and divided into sections of 20 seconds each. Table 2 shows the breakdown of the 20 second segments.

Table 2. Variable message rates by segment

| Segment | Time Period | Data rate/node |
|---|---|---|
| 1st | 1-20 | Pings only |
| 2nd | 21-40 | 5 messages/sec |
| 3rd | 41-60 | Max Rate |
| 4th | 61-80 | 5 messages/sec |
| 5th | 81-100 | Max Rate |

The results of varying the learning rate between 0.1, 0.5, and 1.0 are shown in Figure 7. The data is taken from one node with only 2 Ethernet interfaces therefore the viable actions space is either forward via one of the two interfaces or compute locally, shown as "Forwarding Via Interface 1", "Forwarding Via Interface 2", and "Compute Locally" respectively.

256                    Computer Science & Information Technology (CS & IT)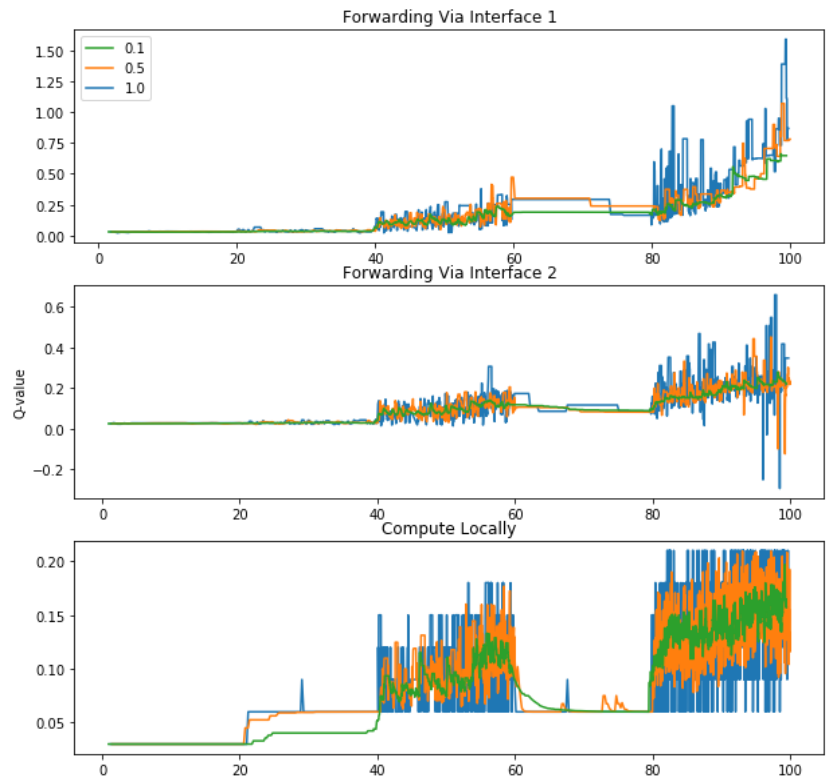

Figure 7. Effect of different learning rates on action space

The results show the increase in variability of the Q-value for higher values of learning rate. Given the Q-value is used to predict the future reward of the a given routing decision, higher variability may not be desirable. For lower values of learning rate, variability is obviously markedly decreased however, the system requires a long time to recover from transition points. A good example of this can be seen at the 60 second mark in the "Compute Locally" section. When the network transitions from high data rate to low data rate the Q-value for computing locally takes approximately 5 seconds to recover. The difference in variability between forwarding and computing is also notable indicating separate learning rates for both processes may be more ideal. From this simple experiment it can be surmised that a learning rate closer to 0.1 is preferable for the forwarding actions while a learning rate closer to 0.5 may be preferable to allow quicker recovery during transition periods. Variable learning rates dependent on network loading could be examined further as an alternate option.

The effect of varying the exploration parameter, ε, was also examined. Epsilon was varied from 0.0, which equates to no exploration and only exploiting the optimal known Q-value, to 0.5 which equates to randomly exploring different actions half of the time. The results are shown in Figure 8.

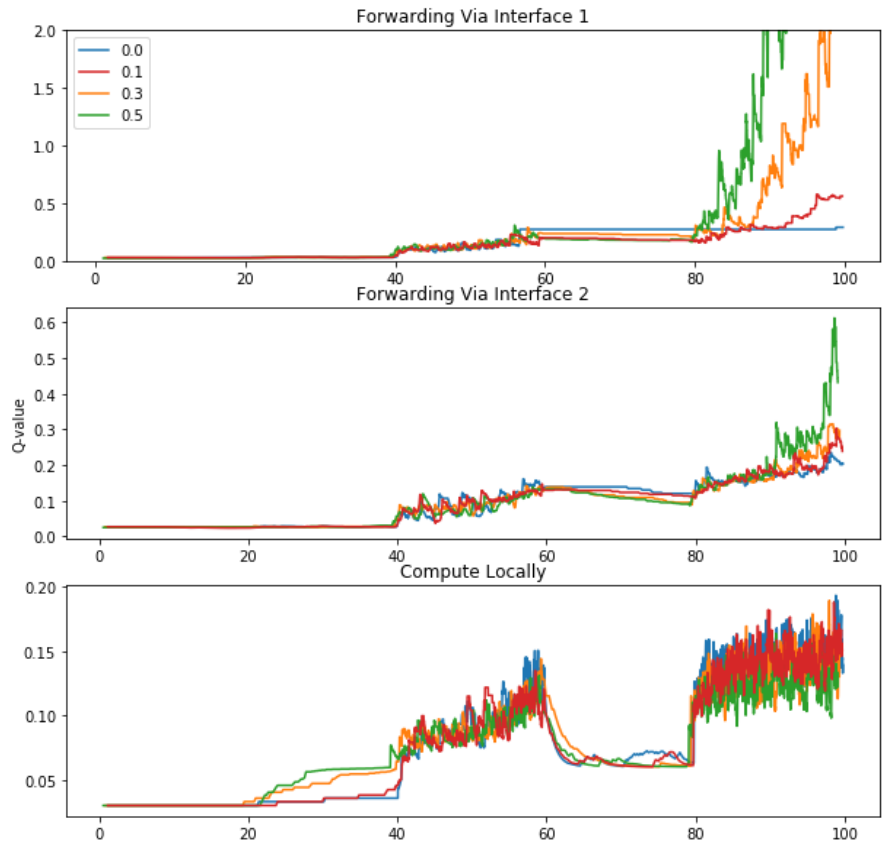

Figure 8. Effect of different ε on action space

The results of varying ε highlight interesting patterns. For the first several sections of the test, the different values have minimal effect on the reward received for a given action, but later in the test as energy reserves begin to deplete there is a large divergence between different values, particularly notable in Interface 1. This is likely due to the cost of forwarding increasing rapidly while energy reserves deplete and randomly selecting forwarding during exploration results in a higher Q-value or lower reward. The option to compute locally seems to benefit from a higher exploration rate. This is likely due, again, to the influence of the energy awareness. As the energy reserves deplete the optimal decision is heavily skewed toward computing locally, however random exploration chooses more messages to be forwarded thereby allowing the compute queue to shrink and lowering the compute locally Q-value. For this particular experiment it appears an ε of 0.1 is ideal as it keeps all options relatively balanced particularly during high data rates and low energy reserves.

## 5. CONCLUSIONS

This paper has presented a variation of Q-routing algorithm with a focus on usability for WSNs. The addition of the option for the algorithm to determine if computational offloading, instead of forwarding, allows optimization of total processing time while not requiring any one node in a multi-hop network to bear the cost of performing the data computation. The result of this addition allows the EAQCO algorithm to optimize for both time and energy simultaneously, while the energy awareness factor alters the action space to account for depleting energy reserves. The results of the simple experiments performed have shown that EAQCO is a viable option when energy, time, and computational capacity are critical factors as is often the case for a WSN.

Additionally, the simplicity of this algorithm allows it to be integrated into other WSN routing protocols particularly other variations of the Q-routing algorithm, further optimizing the capability of the network.

Based on the results of our experiments we intend to continue to expand the action space of EAQCO algorithm in future research to include store-and-carry-forward options in addition to forwarding and computing locally while simultaneously integrating other quality of service metrics such as variable transmission power and link reliability to expand the possible state space. These additional metrics could further increase the utility of the EAQCO for future deployment in WSN.

## REFERENCES

[1] W. Heinzelman, A. Chandrakasan, and H. Balakrishnan, "Energy efficient communication protocol for wireless microsensor networks," in Proceedings of the 33rd Annual Hawaii International Conference on System Sciences, Jan. 2000, pp. 10 pp. vol.2.

[2] C. Ambekar, D. Mehta, and H. Ashar, "OPEGASIS: Opportunistic Power Efficient Gathering in Sensor Information Systems," in 2015 IEEE 9th International Conference on Intelligent Systems and Control (ISCO), Jan. 2015, pp. 1–5.

[3] N. M. Shagari, M. Y. I. Idris, R. B. Salleh, I. Ahmedy, G. Murtaza, and H. A. Shehadeh, "Heterogeneous Energy and Traffic Aware Sleep-Awake Cluster-Based Routing Protocol for Wireless Sensor Network," IEEE Access, vol. 8, pp. 12 232–12 252, 2020.

[4] N. Sadagopan and B. Krishnamachari, "Maximizing Data Extraction in Energy-Limited," International Journal of Distributed Sensor Networks, vol. 1, Apr. 2004.

[5] L. Kong, J.-S. Pan, V. Sn´aˇsel, P.-W. Tsai, and T.-W. Sung, "An energy aware routing protocol for wireless sensor network based on genetic algorithm," Telecommunication Systems, vol. 67, no. 3, pp. 451–463, Mar. 2018.

[6] S. Abolfazli, Z. Sanaei, E. Ahmed, A. Gani, and R. Buyya, "Cloud-Based Augmentation for Mobile Devices: Motivation, Taxonomies, and Open Challenges," IEEE Communications Surveys Tutorials, vol. 16, no. 1, pp. 337–368, 2014.

[7] E. Baccarelli, N. Cordeschi, A. Mei, M. Panella, M. Shojafar, and J. Stefa, "Energy-efficient dynamic traffic offloading and reconfiguration of networked data centers for big data stream mobile computing: review, challenges, and a case study," IEEE Network, vol. 30, no. 2, pp. 54–61, Mar. 2016.

[8] N. Shan, Y. Li, and X. Cui, "A Multilevel Optimization Framework for Computation Offloading in Mobile Edge Computing," Jun. 2020, iSSN: 1024-123X Library Catalog: www.hindawi.comPublisher: Hindawi Volume: 2020.

[9] M. Littman and J. Boyan, "A distributed reinforcement learning scheme for network routing," in In Proceedings of the 1993 International Workshop on Applications of Neural Networks to Telecommunications. Erlbaum, 1993, pp. 45–51.

[10] R. Desai and B. P. Patil, "MANET with Q Routing Protocol," International Journal of Emerging Technologies in Computational and Applied Sciences (IJETCAS), vol. 3, no. 2, pp. 255–262, Feb. 2013.

[11] T. Hendriks, M. Camelo, and S. Latr´e, "Q2-Routing: A Qos-aware Q-Routing algorithm for Wireless Ad Hoc Networks," in 2018 14th International Conference on Wireless and Mobile Computing, Networking and Communications (WiMob), Oct. 2018, pp. 108–115, ISSN: 2160-4886.

[12] S. Khianjoom and W. Usaha, "Anycast Q-routing in wireless sensor networks for healthcare monitoring," in 2014 11th International Conference on Electrical Engineering/Electronics, Computer, Telecommunications and Information Technology (ECTI-CON), May 2014, pp. 1–6.

[13] Xiaolong Xu et al. "Multi-objective computation offloading for Internet of Vehicles in cloud-edge computing," in Wireless Networks 26.3 (Apr. 2020), pp. 1611–1629. ISSN: 1572-8196.doi:10.1007/s11276-019-02127-y.

[14] Z. Hong, W. Chen, H. Huang, S. Guo and Z. Zheng, "Multi-Hop Cooperative Computation Offloading for Industrial IoT–Edge–Cloud Computing Environments," in *IEEE Transactions on Parallel and Distributed Systems*, vol. 30, no. 12, pp. 2759-2774, 1 Dec. 2019, doi: 10.1109/TPDS.2019.2926979.

[15] A. Carroll and G. Heiser, "An Analysis of Power Consumption in a Smartphone," p. 14.
[16] J. Vazifehdan, R. V. Prasad, M. Jacobsson, and I. Niemegeers, "An Analytical Energy Consumption Model for Packet Transfer over Wireless Links," IEEE Communications Letters, vol. 16, no. 1, pp. 30–33, Jan. 2012.
[17] Lora Alliance Technical Committee, "LoraWAN 1.1 Specification," Oct. 2017. [Online]. Available: https://lora-alliance.org/sites/default/files/2018-04/lorawantm specification -v1.1.pdf
[18] "BeagleBoard.org - black." https://beagleboard.org/black